# Influence of Information Support System on ICT Use by Distance Learners in University of Lagos Nigeria

Ugonna Aralu[*]     Airen Adetimirin
Department of Library, Archival and Information Studies, University of Ibadan, Nigeria
Email address: uoaralu@gmail.com

**Abstract**
The use of modern technology in Education is the key to an increased drive for learning which shape learners critical and analytic competencies with respect to disciplinary knowledge. Distance education (DE) is a system of learning driven by computer linked to internet. The flexible nature of DE avail students who are unable to attend full time education due to age, social or religious barriers. However, in Nigeria, the University of Lagos distance learning Institute has its shortfalls traced to poor student support system, which affects the service delivery to students. The study examined Influence of Information Support on ICT use by distance learners. Correlational survey was used in which Information Support System and ICT use was observed. The study population consists of representative students from faculty of Education, Business Administration and Social Sciences make up the study population. Multistage sampling was used with 3% of each population from eight departments, randomly selected to represent sample size of 255 respondents. The findings ranked high Administrative Support, Student Support System. Cybercafé is the most available internet access point to distance learners. The study concludes that Information support on ICT use have impact on distance learners in University of Lagos Nigeria.

## 1. Introduction

Distance education (DE) has always been known for its departure from the conditions in which teaching and learning "naturally" take place. Distance education is anchored on computer mediated technology to develop learner's sense of purpose and general utilization of technological devices, which help shapes critical and analytical competencies with respect to disciplinary knowledge. The perceived "unnaturalness" of distance education has been consistent with the divide between "university proper" and university extension", and with the location of distance programs at the periphery of university life. The flexible nature of distance education, has made it possible for large number of people globally, who have missed university education or have been denied space in university, due to age, social, economic and religious reasons to attend full-time university education. Moore and Kearsley (2008) defined distance education as "planned learning that normally occurs in a different place from teaching, requiring special course designs, instructional techniques, and communication through various technologies, with special organizational and administrative arrangements".

The purpose of distance education is to help bridge the gap in access to higher education. The distance learning students are adults between the ages of 25 and 50 years as described by Moore and Kearsely (2008), packaged to suit adult learners as daily work schedules and engagements in our complex society has made people willing to further their education to enable them to have a job security. The interactive nature of DE, promotes experimentation of new techniques and modern ways of sharing information as contained in UNESCO's strategic objectives in distance education. (UNESCO, 2005). Contacts between the student and institutions are provided through interactive and non-interactive media (Bolt and Wauchope, 2011). It is imperative, therefore, for countries to provide their young people with quality education that would enable them to compete in the global marketplace, as well as, enable them to benefit from the growing socio-cultural global exchanges of the 21[st] century (World Bank, 2011).

Moore and Kearsely (2005) defined information and communication technology ICT as an innovative approach for delivering computer mediated, well-designed, learner-centered and interactive learning environments to anyone, anyplace, anytime by utilizing the internet technologies concerned with instructional design principles. ICT tools comprises of electronic devices which are utilized for information, by teachers and students. Such electronic devices include computer (hardware's and software's), networking devices, telephone, video, multimedia and internet. Application and utilization of these devices converts information, text messages, sound and motion to common electronic form. The purpose of using ICT, in distance learning includes, Access to record and information about school activities fast and timely. It also gives room for unrestricted access to teacher and students in sharing relevant information such as access to course materials and class forum participation development in various fields of study.

## 1.1 Information Support Systems
Information Support System as described by Bialobrezeska and Cohen (2003) are the technologies that generally support an individual's ability to manage and communicate information electronically, and it includes software and systems needed for communication, such as the Internet and e-mail. The information support system comprises of human and non-human resources to guide and facilitate educational transaction through academic social network known as 'social messengers' which runs on the platform of world-wide-web (www). The social messengers is complemented by e-mail, instant messaging, internet phones, video-conferencing, net meeting, weblogs (blogs), and many other systems of communication and chat rooms forums which provide opportunities for students to interact with their tutors and receive feedback targeted at their individualized support (Perrin, 2006).

Information Support Systems (ISS) comprised of search engines which enhances student's easy information retrieval. The availability of Information support system provides the services needed for successful learning environment, which provides adequate information about the program and its outcomes to enable realistic choices by students, robust administrative and technical systems, clear academic expectations and guidance (Kuiper, Volman, and Terwel, 2005). Information support system involves the provision of services and support mechanisms that enhances distance learners/educators on the quality of their research, improve retention and achievement (Brown et al., 2003). Information Support system has a wide range of activities and equipment including all the tools, applications and information, which are available and accessible through computers. It encompasses various forms of information delivery systems such as televisions, radios, newspapers, computers, and the internet (Okwudishu, 2004). The influence of ISS through web based medium of communication has contributed to the development of social presence among the distance learning group (Kehrwald, 2008).

## 1.2 Infrastructural Nature of Information Support System
The infrastructure comprises hardware or equipment, software applications and services associated with ICTs, including the network infrastructure, computing infrastructure, systems and application software, Internet Service Provider (ISP), bandwidth, policy framework and the security infrastructure as Murali (2009) rightly described. The institution, that provides education in open distance learning mode requires a structured network at all its operational nodes and interconnected to each other through a dedicated network so that all student services can be accessed easily by all operational nodes, students and other public. Infrastructural facility is central in achieving the goal of enabling, sustainable and affordable access to ICT (Anderson, 2007). The availability and use of infrastructural facility is however hampered by poor power supply, lack of network infrastructure in particular; high cost of ICT, skills level and lack of enabling policy environment.

According to UNESCO (2003) although government continues to fund education, some degree of fiscal mismanagement or inefficient budget planning leaves little room for improvement of institutional services and operations after the payment of staff salaries and allowances. In addition, UNESCO recommended that 26% of the national budget should be allocated to the education sector. This has never been met by the Nigerian government at any point in time. In 1999, 11% of the Nigerian annual budget was allocated to education; this fell to 6.9% in 2002 and fell further to 1.8% in 2003, but rose to 10.5% in 2004 (Akpotu and Akpochafo,2009). Obviously, the Federal government's budgetary allocation to education has been much less than the recommendation of UNESCO. It is, therefore, not surprising that there has been little corresponding improvement of institutional facilities and infrastructure, thereby greatly reducing access for students seeking higher education in the country.

The university system has been unable to accommodate the ever-growing number of qualified candidates seeking higher education in the country (British Council, 2011). This demand has risen to such high levels that the distance educational institutions have been seriously overstretched with high rate of student's dropout (Akpotu and Akpochafo, 2009). This ever-growing number of students over the years, hinders government's intention to provide facilities and necessary infrastructure for the promotion of ICT at all levels of education due to lack of adequate funding, favourable policies implementation to tackle problems of lack of electricity, insufficient computers, and bandwidth which have adverse effect on use of ICT by distance learners remains unaddressed. Furthermore, the inability of governments to increase funding to the university system has been largely due to financial constraints arising from the slow growth of the nation's economy (Richardson, 2009).

## 2.1 Research Questions and Hypothesis
The study seeks to provide answers to the research questions and to test the hypothesis:
- What is the available Information System Support to distance learners in University of Lagos?
- What are the infrastructural facilities available to distance Learners in University of Lagos?
- What is purpose and frequency of using ICT by distance learners?
- What are the challenges that affect ICT use by distance learners?

Hypothesis: There is no significant relationship between Information Support System and ICT Use by Distance Learners.

## 3. Methodology
### 3.1 Descriptive Survey
The study adopted the descriptive survey design of correlation type. The estimate study population adopted multistage sampling during the selection of students from distance learning Institute University of Lagos.

### 3.2 Study Population
The study population consists of representative students from faculty of education, Business administration and Social Sciences with a total study population 8414. Three faculties participated in the study with strata of eight (8) departments. Three per cent (3%) of each population in all the eight departments was randomly selected, to represent a sample of 255 respondents for this study.

### 3.3 Instruments
The instruments used to collect data for this study was questionnaire and an interview. The research questions were analyzed using statistics, such as mean, standard deviations and variance, while the hypothesis was tested with Pearson correlation coefficient and Analysis of Variance (ANOVA) at 0.05 level of significance.

## 4. Results and Findings

Table 1: Demographic Characteristic of Respondent

| Gender Age and Marital Composition | No | Percentage (%) |
|---|---|---|
| Male | 149 | 58.4 |
| Female | 106 | 41.6 |
| 16-25 years | 63 | 24.7 |
| 26-35 years | 153 | 59.6 |
| 36-45 years | 31 | 12.2 |
| 46-55 years | 9 | 3.5 |
| Single | 161 | 63.1 |
| Married | 86 | 33.7 |
| Divorced | 5 | 2.0 |
| Widowed | 3 | 1.2 |

| Respondent's Faculty | | |
|---|---|---|
| Faculty | No | Percentage (%) |
| Education | 18 | 7.1 |
| Business Administration | 81 | 31.8 |
| Social Sciences | 156 | 61.2 |

### 4.1 What is the available information support system to distance Learners?
The availability of information support system to distance learners in university of Lagos Nigeria was defined by sixteen components and they were measured with a scale of items in order to determine their level of availability. The overall result shows that information support system available to distance learners provides administrative support services of pre-course meeting, internet registration, payment and utilization of learning management system with a mean value of 3.14, ranked highest in the mean score rating. The presence of an established database that contains frequently asked questions, as well as answers to such questions are systematically sorted and filed for future reference. Information support system provides personal and subject tutors, pre-course/post-course announcement of exam results, schedule of online registration and learning co-ordination and special events. The continuous evolution accommodates new learner populations and educational development. The review of course lessons and provision of learning advice by teachers, advisors and learning supervisors has the least mean value of 2.79.

In an oral interview granted to an Information Technology (IT) staff of Distance Learning Institute University of Lagos on his experience with Information Support Systems. He said "Information Support System enhances his presentation in a precise way giving clear directions to the faculty/learners, through the use of e–mails services that enhance communication. Submission of assignments to teachers or appointments and collaborate with the faculty members and students". This response seems to show that there is considerable level of use of communication technology access by distance students. Therefore there is strong evidence, on the use of Information Support System exists.

Table 2: Information Support System

| S/N | Administrative Information Support System | Strongly Disagree (SD) N % | Disagree (D) N % | Agree (A) N % | Strongly Agree (SA) N % | Mean | Std. Dev. |
|---|---|---|---|---|---|---|---|
| 1. | Support pre-course meeting, internet registration, payment and utilization of learning management system | 22  8.6 | 11  4.3 | 132  51.8 | 90  35.3 | 3.14 | .85 |
| 2. | Established team to give services through email and telephone services. | 20  7.8 | 27  10.6 | 118  46.3 | 90  35.3 | 3.09 | .88 |
| 3. | Provides information on enrolment and registration. | 26  10.2 | 22  8.6 | 127  49.8 | 80  31.4 | 3.02 | .90 |
| 4. | Support online communication media for educating the learners of specific subjects, which solve practice problems | 29  11.4 | 22  8.6 | 121  47.5 | 83  32.5 | 3.01 | .93 |
| 5. | Provides an environment which support students, creates commitment and enhances self esteem. | 29  8.6 | 31  12.2 | 125  49.0 | 77  27.8 | 3.01 | .88 |
| 6. | Support communication through telephones, radios, sound track, audiovisual media and television programs | 28  11.0 | 22  8.6 | 134  52.5 | 71  27.8 | 2.97 | .90 |
| 7. | Established database that contains frequently asked questions, as well as answers to such questions, systematically sorted and filed. | 28  11.0 | 29  11.4 | 128  50.2 | 70  27.5 | 2.94 | .91 |
| 8. | Exists to further the goals of a particular institution and serves the need of its learners with its specific context | 32  12.5 | 25  9.8 | 134  52.5 | 64  25.1 | 2.90 | .92 |
| 9. | Integrates multi-media learning activities, review classes and facilitate course teaching assistant's staff. | 32  12.5 | 25  9.8 | 134  52.5 | 64  25.1 | 2.90 | .92 |
| 10 | Support records of learners performances and the administrations, as well as the provision of advise before the course | 36  14.1 | 23  9.0 | 130  51.0 | 66  25.9 | 2.89 | .95 |
| 11. | Information support provides both personal and subject tutors | 34  13.3 | 27  10.6 | 133  52.2 | 61  23.9 | 2.87 | .93 |
| 12. | Support pre-course curriculums which includes primary meeting and training course in using learning system | 30  11.8 | 36  14.1 | 127  49.8 | 62  24.3 | 2.87 | .92 |
| 13. | Supports post-course announcement of exam results, schedules online registration, learning co-ordination and special events. | 45  17.6 | 16  6.3 | 129  50.6 | 65  25.5 | 2.84 | 1.00 |
| 14. | Evolve continuously to accommodate new learner populations, educational developments, economic conditions, technological advances, and finding from research and evaluation. | 38  14.9 | 37  14.5 | 111  43.5 | 69  27.1 | 2.83 | .99 |
| 15 | Supports course during online-offline communication channels | 44  17.3 | 28  11.0 | 116  45.5 | 67  26.3 | 2.81 | 1.01 |
| 16. | Enables review of course lessons and provision of learning advice by teachers, advisors and learning supervisors. | 47  18.4 | 23  9.0 | 122  47.8 | 63  24.7 | 2.79 | 1.02 |

**4.2 What infrastructural facilities are available and frequency of use by distance learners?**

The infrastructural facility available and frequency of use by students was defined by the degree of access to internet facility as elicited in the research survey. The ownership percentage of computers, home, office and school to access internet, level of cybercafé usage and proximity was used to determine the level of communication technology access by distance learners in University of Lagos. The research results indicates the use of cybercafés closer to school very often from a walking distance with mean value of 2.12 value. Also, the Percentage of response on Infrastructural facility enhances student's readiness for learning, including their desire and ability to engage in distance learning and promotes self-development strategies so that student can accept responsibility for developing their own skills has the highest mean value of 3.00. The research results show that people with the same interest to gather and exchange ideas, ask for helps, answers questions, provide and retrieve information were rated with least mean value of 2.89.

Table 3: Available Infrastructure

| S\N | Internet Access | Daily N % | Weekly N% | Monthly N% | Never N% | Mean | Std. Dev. |
|---|---|---|---|---|---|---|---|
| 1 | Cybercafé | 68  26.7 | 59  23.1 | 33  12.9 | 95  37.3 | 2.12 | 1.05 |
| 2 | Home | 142  55.7 | 54  21.2 | 7  2.7 | 52  20.4 | 2.06 | .72 |
| 3 | University/college | 70  27.5 | 47  18.4 | 25  9.8 | 113  44.3 | 1.94 | 1.01 |
| 4 | Office/workplace | 132  51.8 | 30  11.8 | 14  5.5 | 79  31.0 | 1.92 | .80 |
| 5 | Online Databases | 95  37.3 | 41  16.1 | 18  7.1 | 101  39.6 | 1.91 | .91 |
| 6 | School | 71  27.8 | 47  18.4 | 20  7.8 | 117  45.9 | 1.88 | .97 |
| 7 | Public library | 52  20.4 | 34  13.3 | 26  10.2 | 143  56.1 | 1.78 | 1.03 |
| 8 | Friend's/relation's home | 64  25.1 | 35  13.7 | 18  7.1 | 138  54.1 | 1.74 | .95 |
| 9 | Other public internet access point (e.g. public telephone kiosk) | 37  14.5 | 36  14.1 | 19  7.5 | 163  63.9 | 1.65 | .98 |
| 10 | Other (please specify) | 16  6.3 | 11  4.3 | 7  2.7 | 221  86.7 | 1.23 | .66 |

**4.3 What is the purpose of using ICT by distance learners?**

This research question was examined from two angles, that is, from the point of view of the student's functional characteristics (computer/technical skills) and the situational characteristics (Internet access) of distance learners in university of Lagos. Therefore, the dependent variable, the levels of distance learner's computer functional/technical skills has been quantitatively measured by learner's responses to independent variables, such as degree of computer literacy, frequency of computer usage, involvement in social networking, and ownership of e-mail addresses as shown.

Table 4: Purpose of Information and Communication Technology Usage

| S/N | Uses of Information and Communication Technology | Daily N % | Weekly N % | Monthly N % | Never N % | Mean N % | Std. dev. N % |
|---|---|---|---|---|---|---|---|
| 1 | Preparing assignments | 59  23.1 | 64  25.1 | 39  15.3 | 93  36.5 | 2.19 | 1.09 |
| 2 | Email to fellow students/tutors | 81  31.8 | 65  25.5 | 30  11.8 | 79  31.0 | 2.18 | 1.00 |
| 3 | Accessing educational material | 88  34.5 | 54  21.2 | 24  9.4 | 89  34.9 | 2.05 | .97 |
| 4 | Mobile Telephone | 141  55.3 | 41  16.1 | 13  5.1 | 60  23.5 | 2.03 | .78 |
| 5 | Computer | 139  54.5 | 41  16.1 | 11  4.3 | 64  25.1 | 2.00 | .77 |
| 6 | E-books are adequately provided. | 54  21.2 | 44  17.3 | 37  14.5 | 120  47.1 | 1.99 | 1.11 |
| 7 | Television | 120  47.1 | 31  12.2 | 18  7.1 | 86  33.7 | 1.93 | .86 |
| 8 | Preparing for travel/holiday | 39  15.3 | 35  13.7 | 36  14.1 | 145  56.9 | 1.85 | 1.12 |
| 9 | Digital Video Disk/CD ROM players are available. | 54  21.2 | 43  16.9 | 21  8.2 | 137  53.7 | 1.80 | 1.00 |
| 10 | Flash drives/External Hard drives are adequately provided. | 50  19.6 | 48  18.8 | 16  6.3 | 141  55.3 | 1.76 | .97 |
| 11 | Software is sufficiently provided. | 41  16.1 | 30  11.8 | 27  10.6 | 157  61.6 | 1.71 | 1.04 |
| 12 | Video conferencing | 38  14.9 | 37  14.5 | 20  7.8 | 160  62.7 | 1.67 | .99 |
| 13 | Other (please specify) | 15  5.9 | 17  6.7 | 8  3.1 | 215  84.3 | 1.29 | .73 |

**4.4 What are the challenges that affect ICT use by distance learners?**
The challenges that affect ICT use by distance learners was examined quantitatively, with additional qualitatively research as to provide more clarity to research findings. The instructional audio tapes ranked highest in the mean score rating, with a mean value of 2.04 on mobile phones and mean value of 2.03 on computers, as shown. The table shows that distance education students surveyed had a computer ownership rate of 12% with home monthly Internet access of 2.7%. Respondents also indicated 11.8% smart phone readily available and had access to the Internet with their smart phones. In addition, 26.7% of distance education students surveyed use cybercafés on daily basis to meet their academic and social needs.

In an oral interview with the IT staff, who indicated inadequate Infrastructural facilities, slow human capital development, finance and a host of other factors that contributes to poor facilitation of study, teaching and learning process in terms of information dissemination, record keeping, poor quality and design of course materials, and general administration. "They sometimes fail to get the necessary information about studies such as: deadline for submission of term papers, meetings with facilitators, poor student's record keeping, et.cetera" (Staff)

In view of the statement it can be deduced that unnecessary delays are caused in terms of information dissemination as a result of absence of this facilities. He also narrated challenges encountered using internet with communication technology to facilitate communication with students. Some of these challenges cited include: High bandwidth costs, limited access to the Internet and technology as majority of the students, especially those in commuting from home/outside the state, do not have access to ICT facilities. High illiteracy rates which influence student's ability to use these facilities when available. Resistance to organizational change as organizations should change the way they provide services and do their work, the need for student to adapt. However, in most cases, Staff resists changes, and stick to old traditional methods, which makes the implementation of ICT as tools very difficult.

Computer literacy is also a factor as many organizations have computer systems in offices, but they are not used by staff due to their inability to use them. Lack of knowledge and information management competencies, there is very little that a person can do without knowledge (or know-how). Knowledge is power, and a staff member with little or no knowledge (of ICT) will never perform as well as someone who knows all about ICT.

**4.5 Respondents Challenges on ICT**
There was significant relationship between information system support and ICT use by distance learners as shown in table 4.4.6 at * Sig .05 level (r = .225*, N= 255, P < .05).

**4.6 Suggestions and Recommendations**
In the strive to promote academic excellence by institutions of higher learning to effectively operate distance education programs, they are faced with numerous problems that hinders the adequate implementation to meet the policy as contained in the Federal Republic of Nigeria (2004) through its national policy on education detailed that the goal of distance education should be to:
(1) Provide access to excellent education and equity in educational opportunities for those who otherwise would have been denied.

(2) Meet special needs of employers by mounting special certificate courses for their employees at their work place.
(3) Encourage internationalization especially of tertiary education curricula.
(4) Restructure the effect of internal and external brain drain in tertiary institutions by utilizing Nigerian experts as teachers regardless of their locations or places of work.
The federal government from the aforementioned goals should implement that for distance learning education to make optimum contribution to national development, information and communication technologies is an essential ingredient to foster its implementation.

## 5. Conclusion

In view of the research finding the nature of this study contains effect of information support system on ICT use by distance learners. At a broad management level, this study calls for effective policies to make a balanced investment in distance education programs and provide resources needed to effectively implement the use, integration and diffusion of ICT in distance learning rather than paying eye service. The impact was found to be relevant in administrative roles which support access to distance learners. ICT provides solutions to specific problem of system of administration, decision-making in the administration of distance learning as it guarantees effective administrative practices and competency of administrators and learners.

**References**
Moore, M. & Kearsley, G.2008. Distance Education: A Systems View. 2nd ed. Thomson Wadsworth, 8-23
Moore, M. G., & Kearsley, G. 2005. Distance education: A systems view (2nd ed.). Belmont, CA: Wadsworth Publishing Company.
Bolt, S., Kerr, R. & Wauchope, V. 2011. 'Using video analysis software to create innovative teacher professional development', paper presented to the scientific meeting of the International Conference on Information Communication Technologies in Education, Rhodes.
World Bank 2011. Tertiary Education. Retrieved from: http://www.worldbank/education/tertiary; http://go.worldbank.org/HBEGA0G2P0
Bialobrzeska, M. & Cohen, S. 2005. Managing ICTs in South African schools: Aguide for school principals. Braamfontein: South African Institute for Distance Education.
Perrin, D. G. 2006. It's a number game. International Journal of Instructional Technology & Distance Learning, 3(11), 1. Retrieved from http://www.itdl.org
Kuiper, E.; Volman, M.; & Terwel, J. 2005. The Web as an Information Resource in K-12 Education: Strategies for supporting students in searching & processing information.
Review of Educational Research, 75(3):285-328. ERIC (EJ737303).
Brown, C., Murphy, T.J., & Nanny, M. 2003. Turning techno-savvy into info-savvy: Authentically integrating information literacy into the college curriculum. The Journal of Academic Librarianship, 29(6), 386-398.
Okwudishu, C.O. 2005. ICTs and digital divide: Implications for open learning and distance education in Nigeria Journal of Education in Developing Areas (JEDA) (13) 202-210
Kehrwald, B. 2008. Understanding social presence in text-based online learning environments, Distance Education, 29(1), 89-106.
Murali M Rao, 2009. 'Web-enabled User Support Services System in Distance Learning',Proceedings of International Conference on Interaction Sciences: Information
Technology, Culture and Human(ICIS 2009), The ACM International Conference Proceeding, Vol I, ISBN 978-1-60558-710-3,Seoul, Korea, pp 86-90.
Anderson, P. 2007. "Ideas, technologies and implications for education. JISC Technology and Standards Watch". http://www.jisc.ac.uk/media/documents/techwatch/tsw0701b.pdf.
British Council 2011. Cross-Border Higher Education in Nigeria – Strategic Partnership and Alliances – Prime Minister's Initiative. Retrieved from: http://www.britishcouncil.org/learning-pmi2-policy-dialogues-tne.htm
Akpotu, N. & Akpochafo, W. 2009. An Analysis of Factors Influencing the Upsurge of Private Universities in Nigeria. Journal of Social Science. 18 (1) 21-27
Richardson, A. 2009. Crossing the Chasm- introducing Flexible Learning into Botswana Technical Education Programmes: From Policy to Action. International Review of Research in Open and Distance Learning. Vol. 10 (4), 1-15
UNESCO (2003). Feasibility Study on the Development of a Virtual Library. Institution of Higher Education in Nigeria.
Federal Ministry of Education (FME) 2007. Education Reform Act. Arrangement of parts (Education Sector Reform Bill).